\documentclass{moriond}
\usepackage{orcidlink}

\def\be{\begin{equation}}
\def\ee{\end{equation}}
\def\bea{\begin{eqnarray}}
\def\eea{\end{eqnarray}}

\newcommand{\Photo}{}

\begin{document}
\vspace*{4cm}
\title{Search for CBCs with SSM Components in Data from The First Part 
of LVK Fourth Observing Run}

\author{I. Bentara \orcidlink{0009-0000-5074-839X} \\
On behalf of the LIGO-Virgo-KAGRA collaboration}

\address{Université Claude Bernard Lyon 1, CNRS, IP2I Lyon \\
IN2P3, UMR 5822, F-69622 Villeurbanne, France}

\maketitle\abstracts{
    Star evolution models predict the lightest compact objects in 
    the universe to have a mass greater than the mass of the sun. 
    Nonetheless, some alternative theories consider scenarios that 
    could lead to the formation of Sub-Solar Mass (SSM) compact objects, 
    such as Primordial Black Holes (PBHs). 
    The LIGO-Virgo-KAGRA Collaboration (LVK) has performed a search for 
    Gravitational Wave (GW) signals emitted by Compact Binary Coalescences 
    (CBCs) including at least one SSM component in the first part of their 
    fourth observing run (O4a), and reports no statistically significant 
    candidate. 
    This absence of detection sets upper limits on the merger rate of
    CBCs including a SSM Black Hole (BH), which can then be used to
    constrain 
    PBH formation models and the fraction of Dark Matter (DM) in PBHs. 
    For PBH binaries forming at late times, the fraction of DM in PBHs is
    obtained to be $\leq 1$ for masses above 0.9 $M_{\odot}$ for 
    monochromatic mass functions. 
    In the early-formation scenario, this fraction is limited to be 
    $\leq 7$\% at $1~M_\odot$, and $\leq 40$\% at 0.35 $M_\odot$.
       }

\section{Introduction}

Since the first GW detection in 2015~\cite{LIGOScientific:2016aoc}, 
 the LVK~\cite{KAGRA:2013rdx} collaboration has identified over 200 
 significant GW candidates~\cite{LIGOScientific:2025slb,LIGOScientific:2025hdt,LIGOScientific:2025yae} 
 from CBC searches with component masses above $1~M_{\odot}$. 
These observations are consistent with stellar evolution models, in which
 BHs should be heavier than neutron stars, whose masses are expected near
 the Chandrasekhar limit of $1.4~M_{\odot}$~\cite{Singh:2020wiq}.
However, alternative formation channels predict SSM compact objects,
 potentially of PBH origin.
While LIGO~\cite{LIGOScientific:2014pky} and 
 Virgo~\cite{VIRGO:2014yos} detectors are primarily sensitive to 
 stellar-mass mergers, they can also probe lighter systems, enabling 
 such searches.

These proceedings summarize recent LVK searches for SSM compact
 objects~\cite{LVK:2026inprep}, focusing on PBH scenarios and the constraints
 on their abundance from non-detections.

\section{O4a SSM search}
\subsection{Method and results}

Three independent search pipelines --~GstLAL~\cite{Cannon:2020qnf}, 
 MBTA~\cite{Alléné_2025} and PyCBC~\cite{Davies:2020tsx,pycbcsoft}~-- 
 analyzed data from O4a, using only data from the two LIGO 
 detectors~\footnote{
    The analysis also includes data from the preceding engineering run,
     restricted to periods when both LIGO detectors were operating
     coincidentally.
    }, 
 as Virgo was not observing.
The three pipelines use matched filtering, correlating the data with their
 own collections of CBC waveform models, referred to as 
 \textit{template banks}, weighted by the detector noise. 
Despite differences in their construction~\cite{LVK:2026inprep}, these
 template banks span the same parameter space.
They target CBCs with redshifted masses $(1+z)m_1 \in [0.2,10.0]\,M_\odot$
 and $(1+z)m_2 \in [0.2,1.0]\,M_\odot$, and mass ratio
 $m_2/m_1 \in [0.1,1.0]~(m_2 < m_1)$. 
Spins are assumed aligned or anti-aligned with the orbital angular momentum, 
 with magnitudes $\chi_{1,2} \leq 0.9$ for $(1+z)m_{1,2} \geq 0.5\,M_\odot$ 
 and $\chi_{1,2} \leq 0.1$ for $(1+z)m_{1,2} \leq 0.5\,M_\odot$. 
This parameter space matches that of the LVK SSM analysis of the third 
 observing run (O3)~\cite{LIGOScientific:2021job,LVK:2022ydq}.

\begin{figure}
\centerline{\includegraphics[width=1.\linewidth]{
       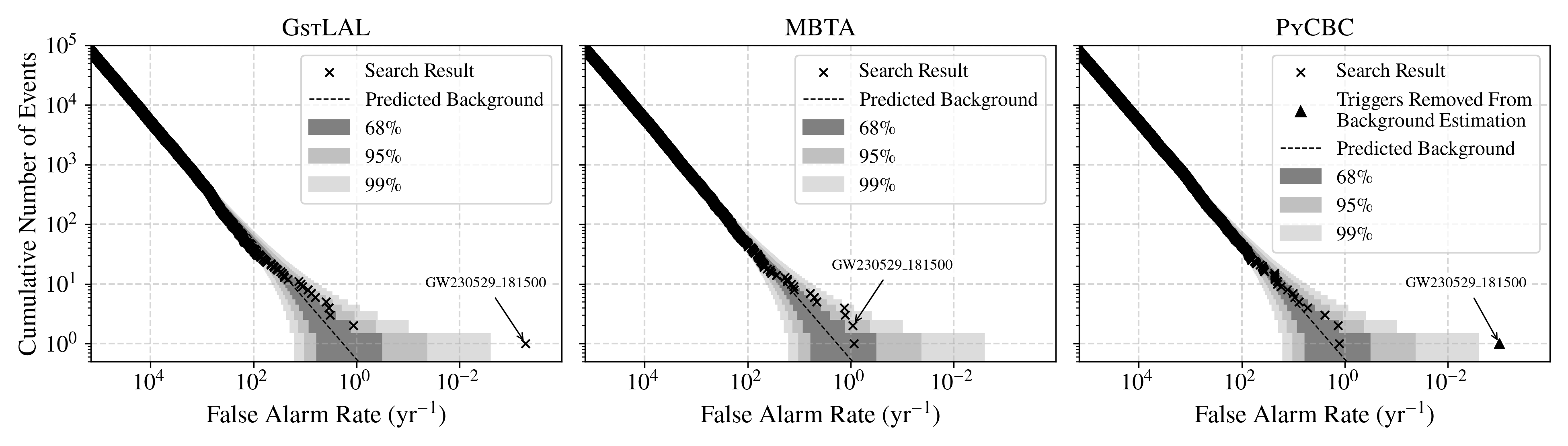}}
\caption[]{Figure reproduced from~\cite{LVK:2026inprep}.
    Cumulative number of SSM candidate events as a function of false-alarm
     rate (FAR) in yr$^{-1}$, for GstLAL (left), MBTA (middle), and PyCBC 
     (right). 
    The dashed curve shows the expected background, with shaded 68\%, 95\%, 
     and 99\% Poisson confidence intervals for O4a. Observed counts are shown
     as black crosses. 
    The most significant trigger,
     GW230529\_181500~\cite{LIGOScientific:2024elc}, has a secondary mass 
     above $1\,M_\odot$ and is therefore not an SSM candidate. 
    In PyCBC, it is excluded (black triangle) due to the removal of events
     with FAR $\leq 1/100\,\mathrm{yr}$ for background
      estimation~\cite{Nitz:2018imz}. 
    No significant deviation from background is observed.
}
\label{fig:far_plot}
\end{figure}

No significant GW transients consistent with SSM candidates are 
 identified by GstLAL, MBTA, or PyCBC.
Figure~\ref{fig:far_plot} shows the GstLAL (left), MBTA (middle), and PyCBC
 (right) results, presenting the number of events above a given 
false-alarm rate (FAR) threshold as a function of that threshold.
The FAR measures the expected rate at which noise fluctuations produce events
 with equal or greater significance.
The observed triggers are consistent with a Poisson distribution of noise 
 background events, with the exception of a single outlier corresponding 
 to GW230529\_181500~\cite{LIGOScientific:2024elc}. 
This event, the most significant trigger in both GstLAL and PyCBC, is a 
single-detector candidate observed at LIGO Livingston and identified in the
 regular LVK super-solar search with a higher 
 significance~\cite{LIGOScientific:2024elc}.
Its chirp mass $\mathcal{M}=\frac{(m_1 m_2)^{3/5}}{(m_1 + m_2)^{1/5}}$ lies
 within the SSM search parameter space, explaining its recovery in these 
 analyses. 
However, its inferred secondary mass exceeds one solar mass, excluding an
 SSM interpretation~\cite{LIGOScientific:2024elc}.

\subsection{Sensitivity, rate and constraints}

\begin{figure}
\begin{minipage}{0.46\linewidth}
    \begin{minipage}{\linewidth}
        \centerline{\includegraphics[width=0.978\linewidth, trim=0.6cm 19.6cm 0.5cm 2cm, clip]{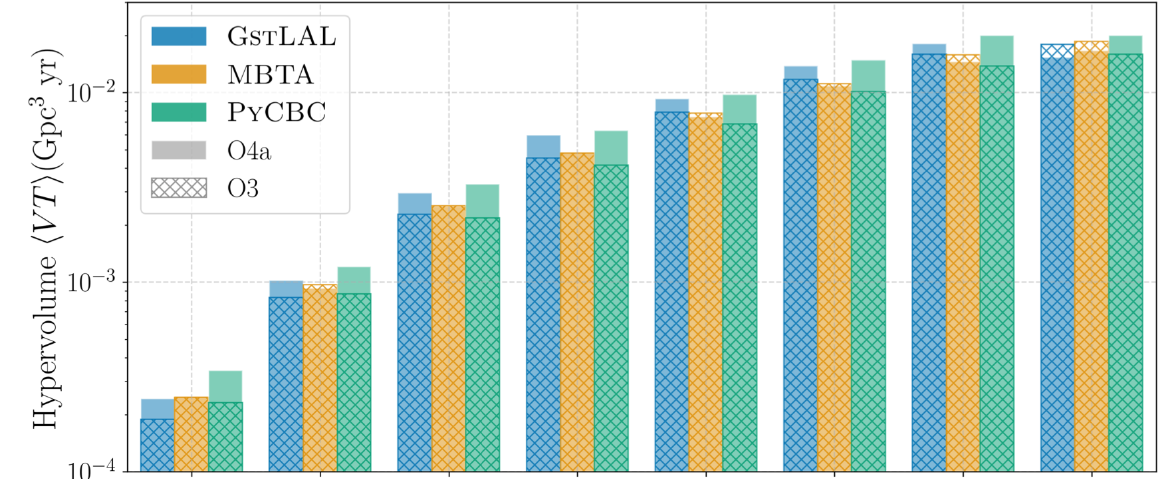}}
    \end{minipage}
    \vspace{0.1cm}
    \begin{minipage}{\linewidth}
        \centerline{\includegraphics[width=0.97\linewidth, trim=0.5cm 18cm 0.5cm 2cm, clip]{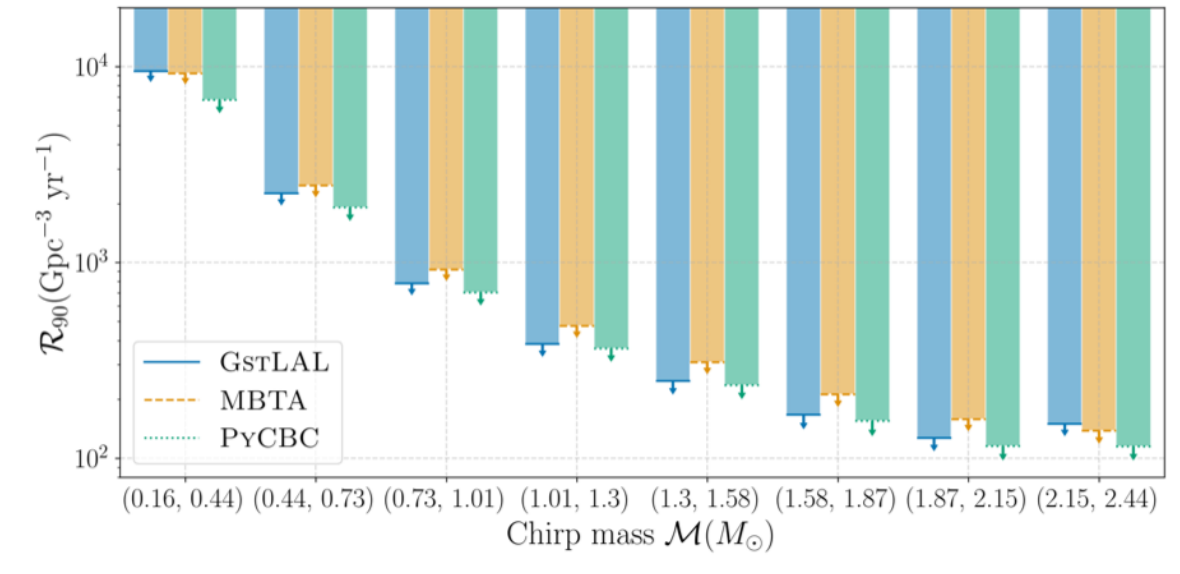}}
    \end{minipage}
\end{minipage}
\begin{minipage}{0.47\linewidth}
    \centerline{\includegraphics[width=0.95\linewidth]{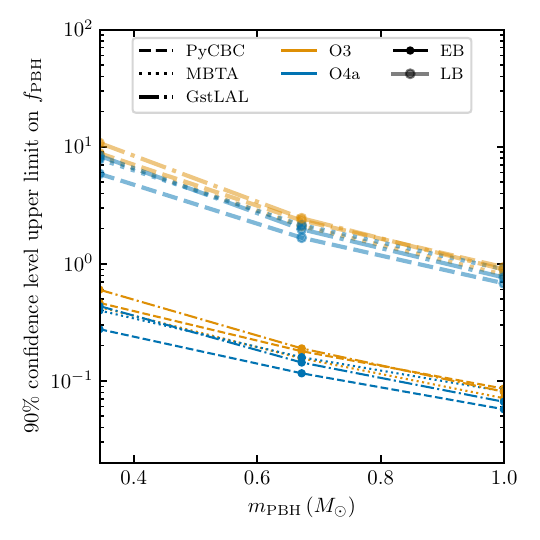}}
\end{minipage}
\caption[]{
    Left: Sensitive spacetime volume $\langle VT \rangle$ (top) and derived
     90\% confidence upper limits on the merger rate (bottom) as functions of
     chirp mass. 
    Right: Upper limits on the PBH DM fraction $f_{\mathrm{PBH}}$ as a
     function of PBH mass, derived from O4a data assuming early and late 
     binary formation scenarios, reproduced from~\cite{LVK:2026inprep}.
}
\label{fig:constraints}
\end{figure}

The search sensitivity is estimated via simulation campaigns and quantified
 by the sensitive spacetime volume $\langle VT \rangle$, which depends on 
 the assumed mass and spin distributions~\cite{Essick:2025zed}. 
Two fiducial populations are considered; results are shown here for BBHs
 with at least one SSM component (see~\cite{LVK:2026inprep} for BNS, 
 including tidal effects). 
Results as function of the chirp mass are shown in the top-left panel of
 Fig.~\ref{fig:constraints}.
The sensitivity achieved in O4a is slightly better although comparable to
 that of O3, as improvements in detector performance are largely offset by 
 the shorter observing time in O4a~\cite{LVK:2026inprep}.

Building on the sensitivity estimates, these results are translated into
 constraints on the merger rates of CBCs with an SSM component.
Assuming a Poisson distribution of triggers, the loudest-event
 formalism~\cite{Biswas:2007ni} is used to derive 90\% confidence upper 
 limits on the merger rate of BBHs with an SSM component depicted in the 
 bottom-left panel of Fig.~\ref{fig:constraints}. 
Given the absence of significant SSM candidates, the rate limits are computed
 as $\mathcal{R}_{90,i} = 2.3 / \langle VT \rangle_i$ in each mass bin. 
The resulting limits show a marginal improvement over O3, reflecting the
 similar $\langle VT \rangle$ observed in O4a~\cite{LVK:2026inprep}.
The merger-rate upper limits are then used to derive constraints on the PBH
 DM fraction $f_{\mathrm{PBH}}$, noting that the mapping between 
 $f_{\mathrm{PBH}}$ and merger rate is model-dependent due to uncertainties 
 in PBH mass distributions and clustering~\cite{LVK:2026inprep}. 
Two formation channels—early binaries (EB) and late binaries (LB)—are
 considered, and limits are derived assuming a monochromatic PBH mass 
 function. 
The resulting constraints from O4a are found to be consistent across
 pipelines, with a slight improvement compared to O3~\cite{LVK:2026inprep}, 
as shown in the right panel of Fig.~\ref{fig:constraints}. 
For PBH binaries forming at late times, the fraction of DM in PBHs is 
 obtained to be $\leq 1$ for masses above 0.9 $M_{\odot}$.
In the early-formation scenario, this fraction is limited to be $\leq 7$\%
 at $1~M_\odot$, and $\leq 40$\% at 0.35 $M_\odot$~\cite{LVK:2026inprep}.

\section{Conclusion}

A search for compact binary coalescences (CBCs) with at least one SSM
 component has been performed using data from the first part of the fourth 
 LVK observing run. 
No significant candidates are identified, and all observed triggers are found
 to be consistent with the expected background, with the exception of 
 GW230529\_181500, whose properties exclude an SSM interpretation.
Upper limits on the merger rates are derived from these non-detections, 
 leading to constraints on the abundance of PBHs as a DM component. 
The resulting bounds are comparable to those from previous observing runs, 
 with slight improvements. 
Future observations with increased sensitivity and duration are expected
 to further improve these limits.

\section*{Acknowledgments} 

This material is based upon work supported by NSF’s LIGO Laboratory 
which is a major facility fully funded by the National Science Foundation, 
as well as the Science and Technology Facilities Council (STFC) of the United Kingdom, 
the Max-Planck-Society (MPS), and the State of Niedersachsen/Germany for 
support of the construction of Advanced LIGO and construction 
and operation of the GEO600 detector. 
Additional support for Advanced LIGO was provided by the Australian Research Council. 
Virgo is funded, through the European Gravitational Observatory (EGO), 
by the French Centre National de Recherche Scientifique (CNRS), 
the Italian Istituto Nazionale di Fisica Nucleare (INFN) and the Dutch Nikhef, 
with contributions by institutions from Belgium, Germany, Greece, Hungary, 
Ireland, Japan, Monaco, Poland, Portugal, Spain. KAGRA is supported by Ministry of Education, 
Culture, Sports, Science and Technology (MEXT), Japan Society for the Promotion 
of Science (JSPS) in Japan; National Research Foundation (NRF) and Ministry of Science 
and ICT (MSIT) in Korea; Academia Sinica (AS) and National Science and Technology 
Council (NSTC) in Taiwan. 

\section*{References}
\bibliography{moriond}

\end{document}